\documentstyle[twocolumn,epsf]{jpsj}
\def\runtitle{ 
Mechanism of confinement in low-dimensional organic conductors
}
\def\runauthor{
Y. Suzumura, M. Tsuchiizu 
}

\def\td{t_{\rm d}}
\def\tperp{t_{\perp}}

\def\dx{{\rm d}x\,}
\def\dl{{d \over{dl}} \,}
\def\e{{\rm e}}

\def\virg{\;\;,}

\def\vf{v_{\rm F}}
\def\kf{k_{\rm F}}

\def\eff{{\rm eff}}
\def\eff0{{\rm eff,0}}

\def\ggs{\buildrel\textstyle > \over {\hbox{\raise0.2ex\hbox{$\sim$}}}}
\def\lls{\buildrel\textstyle < \over {\hbox{\raise0.2ex\hbox{$\sim$}}}}
\def\gsim{\,\lower0.75ex\hbox{$\ggs$}\,}
\def\lsim{\,\lower0.75ex\hbox{$\lls$}\,}
\def\para{\parallel}

\def\ttperp{\tilde{t}_\perp}

\def\Kc   {K_{\rm C}}
\def\Ks   {K_{\rm S}}

\def\gncp {G_{\nu + ,{\rm C} +}}
\def\gncm {G_{\nu + ,{\rm C} -}}
\def\gnc  {G_{\nu + ,{\rm C} p}}

\def\gnsp {G_{\nu + ,{\rm S} +}}
\def\gnsm {G_{\nu + ,{\rm S} -}}
\def\gns  {G_{\nu + ,{\rm S} p}}

\def\gcs  {G_{{\rm C} p ,{\rm S} p'}}
\def\gcsp {G_{{\rm C} p ,{\rm S} +}}
\def\gcsm {G_{{\rm C} p ,{\rm S} -}}

\def\gcps {G_{{\rm C} + ,{\rm S} p}}
\def\gcms {G_{{\rm C} - ,{\rm S} p}}
\def\gcpsp{G_{{\rm C} + ,{\rm S} +}}
\def\gcpsm{G_{{\rm C} + ,{\rm S} -}}

\def\grc  {G_{\rho +, {\rm C} p}}
\def\grcp {G_{\rho +, {\rm C} +}}

\def\grs  {G_{\rho +, {\rm S} p}}
\def\grsd {G_{\rho +, {\rm S} p'}}

\def\gpc  {G_{\sigma +, {\rm C} p}}
\def\gpcp {G_{\sigma +, {\rm C} +}}

\def\gps  {G_{\sigma +, {\rm S} p}}
\def\gpsd {G_{\sigma +, {\rm S} p'}}

\def\tt{\tilde{t}}

\def\cite #1{[\citen{#1}]}

\def\jo #1#2#3#4{#1 {\bf #2} (#3) #4}  

\def\PRB{Phys.\ Rev.\ B}

\def\PTP{Prog.\ Theor.\ Phys.}

\def\EPL{Europhys.\ Lett}

\def\EPJB{Eur.\ Phys.\ J.\ B}

\title
{
Mechanism of confinement in low-dimensional organic conductors
 }

\author{
Y. Suzumura$^{a,b}$,  M. Tsuchiizu$^a$
}

\inst{
$^{a}$Department of Physics, Nagoya University, Nagoya 464-8602, Japan
\\
$^{b}$CREST, Japan Science and Technology Corporation (JST) 
}

\recdate{\hspace{3cm}}

\abst{
 Confinement-deconfinement transition in quarter-filled 
  two-coupled chains comprising  dimerization,  
    repulsive interactions and interchain hopping has been demonstrated 
   by applying the renormalization group method to 
     the bosonized Hamiltonian. 
 The  confinement given by   
   the irrelevant interchain hopping  occurs    
    with  increasing umklapp scattering which is induced by 
 the dimerization leading to  effectively half-filling.   
  It is shown  that the transition originates in 
   a competition between a charge gap and the renormalized 
 interchain hopping. 
}

\kword{
A. Organic compounds, D. Electronic structure, 
 D. Electrical Properties
}

\begin{document}
\maketitle

\section{Introduction}
 In low-dimensional organic conductors,
   repulsive  interactions  play an important role 
   for electronic states with a gap or a pseudo gap. 
  The  anisotropy in electric conductivity 
  is enhanced by   interactions  since the induced  pseudo  gap 
       around the Fermi surface 
  of a single chain precludes electrons from  hopping between chains
  \cite{Bourbonnais}.
 There are several arguments as to whether or not the electrons 
  are confined to a chain 
    by the repulsive interaction.
 Away from half-filling,  the confinement needs 
    a  large magnitude of the  interaction  
      even for the small limit of the interchain hopping
  \cite{Fabrizio} 
 since  the effect of the interchain hopping is much larger 
  than that of the intrachain interaction. 
 However, in the case of half-filling,   
  the  electrons can be confined by 
  the interaction  with a moderate strength   
    due to umklapp scattering which  induces the charge gap
  \cite{Bourbonnais_U,Suzumura,Tsuchiizu}. 

     Bechgaard salts of  organic conductors, TMTSF and TMTTF,  
      can be regarded as  effectively half-filling 
          due to dimerization
  \cite{Jerome,Ishiguro}. 
 The  optical experiments  
  have shown the finite Drude weight  for  the TMTSF salts 
  but not  for the TMTTF  salts  
 although  the correlation gap   exists in both salts 
 \cite{Vescoli}. 
   This indicates    
 the  transition  from an insulating state with the  electrons
  confined to chains to a metallic state with   deconfined electrons
   when the correlation gap becomes  larger than the interchain hopping
  \cite{Moser,Schwartz}. 
 In the present study, 
 such a  transition 
  is elucidated by applying  the  renormalization group (RG) method 
  to a model of quarter-filled two-coupled chains with dimerization.

\section{Formulation}
 We consider  quarter-filled two-coupled chains given by 
\begin{eqnarray}
{\cal  H} &=& - \sum_{j,\sigma,l}
   \left[ t + (-1)^{j} \td\right] 
   \left(  c_{j \sigma l}^{\dagger} 
           c_{j+1 \sigma l} + \mbox{\rm h.c.} \right)
\nonumber \\ &&{}
- 2 t_{\perp} \sum_{j,\sigma} 
\left(
   c_{j \sigma 1}^{\dagger} c_{j \sigma 2}
 + \mbox{\rm h.c.} \right)
+ U \sum_{j,l} n_{j \uparrow l} \, n_{j \downarrow l}  ,
\label{eq:H}
\end{eqnarray}
 where $t$, $\tperp$, $\td$ and $U$ denote energies for 
   the intrachain hopping, the interchain hopping, dimerization 
 and on-site repulsion, respectively. 
 $n_{j \sigma l} = c_{j \sigma l}^\dagger c_{j \sigma l}$. The quantity 
  $c_{j \sigma l}$ denotes the annihilation operator of the electron
   at the $j$-th site
  of the $l$-th chain ($l=$1, 2) with spin 
  $\sigma$($=\uparrow,\downarrow$).
We use the Fourier transform, 
$c_{j \sigma l} = N^{-1/2} \Sigma_{k} c_{k \sigma l} \exp[i kja]$
  with the total number of sites $N$ and the lattice constant $a$.
 First, the $\td$-term is diagonalized to obtain two bands 
 in the reduced zone,   $- \pi /2a < k < \pi/2a$, 
 and the lower band becomes effectively half-filled, 
    which band is described with  fermion operators, $d_{k \sigma l}$, 
    and   is examined in the present study.
 Next, diagonalizing the $t_{\perp}$-term by  
  $a_{k\sigma \pm} = ( \mp d_{k\sigma 1} + d_{k\sigma 2} )/\sqrt{2}$,
  the kinetic term is  written as
   $ {\cal H}_K^d = 
     \sum_{k,\sigma,\zeta} $ 
    $\varepsilon(k,\zeta) $ 
     $ a_{k \sigma \zeta}^\dagger a_{k \sigma \zeta}$
    $(\zeta=\pm)$
 with $\varepsilon(k,\pm)
  = -2 [t^2 \cos^2 ka + \td^2 \sin^2 ka]^{1/2} \pm 2\tperp $.
Thus we have the following effective Hamiltonian \cite{Tsuchiizu_SCES}. 
 The kinetic energy 
   with  the linearized dispersion around the Fermi surfaces,
   $k_{{\rm F} \pm} = \kf \mp \tperp/\vf$, is expressed as 
 ${\cal H}_K^d = 
   \sum_{k,p,\sigma,\zeta} $ $\vf (pk-k_{{\rm F}\zeta}) \,
     a_{k p \sigma \zeta}^\dagger \, a_{k p \sigma \zeta}$
      with  $p (=+,-)$ denoting 
       right moving (left moving) electrons and 
   $\vf=\sqrt{2}ta$ $[1-(\td/t)^2]$ $/[1+(\td/t)^2]^{1/2}$, 
 in which 
  the  $\tperp$-dependence of the velocity is discarded.  
 Coupling constants of interactions corresponding to  
    forward scattering with the same and  opposite  directions 
  ($g_{4}$ and $g_{2}$),
      backward  scattering ($g_1$) and 
          Umklapp scattering ($g_3$) are given by 
  $g_{1^\perp} =  g_{2^\perp} = g_{4^\perp} =  U  a$,
    $g_3 \propto Ua (\td/t)$ 
    and $g_{1^\para} = g_{2^\para} = g_{4^\para}= 0$ 
  where $\para$ and $\perp$ denote interactions for the same spin and 
   opposite spin.

Applying  the  bosonization  method to electrons  around the  
  new Fermi points, 
  we introduce  Bose fields of  phase variables,  
  $\theta_{\rho +}$ and  $\theta_{\sigma +}$
  ($\theta_{{\rm C}+}$ and $\theta_{{\rm S}+}$)
\cite{Suzumura,Tsuchiizu},  
  which  express   fluctuations for 
  the total (transverse) charge density and 
  spin density,  respectively
\cite{Suzumura_PTP}.
The commutation relation with  conjugate phase variables is given by  
$ [\theta_{\nu +}(x),\theta_{\nu'-}(x')]_
   = i \pi \delta_{\nu,\nu'} {\rm sgn}(x-x')$ .
 In terms of  these phase variables,
 our   Hamiltonian is given by  
\begin{eqnarray}
{\cal H} &=& 
\sum_{\nu = \rho,\sigma,{\rm C},{\rm S}} 
\frac{v_\nu}{4\pi} \int \dx
 \left[
   \frac{1}{K_\nu} \left(\partial \theta_{\nu +} \right)^2
         +  K_\nu  \left(\partial \theta_{\nu -} \right)^2
 \right]
\nonumber \\
&& + 
    \frac{g_\rho}{4\pi^2 \alpha^2} \int \dx 
    \left[
        \cos ( \sqrt{2}\theta_{{\rm C +}} - 8 \tperp x/\vf ) 
\right. \nonumber \\ && \left. \hspace{0.5cm}
      + \cos \sqrt{2} \theta_{{\rm C -}}  
    \right]
    \left[
        \cos \sqrt{2} \theta_{{\rm S}}
      - \cos \sqrt{2} \theta_{{\rm S -}}
    \right] \nonumber \\
&&+
    \frac{g_\sigma}{4\pi^2 \alpha^2} \int \dx
    \left[ 
        \cos ( \sqrt{2}\theta_{{\rm C +}} - 8 \tperp x/\vf ) 
\right. \nonumber \\ && \left. \hspace{0.5cm}
      - \cos \sqrt{2} \theta_{{\rm C -}}  
    \right] 
    \left[
        \cos \sqrt{2} \theta_{{\rm S +}} 
             + \cos \sqrt{2} \theta_{{\rm S -}}
    \right] \nonumber \\
&&+ 
    \frac{g_{1\perp}}{2\pi^2 \alpha^2}  \int \dx
    \cos \sqrt{2} \theta_{\sigma +} 
    \left[
        \cos ( \sqrt{2}\theta_{{\rm C +}} - 8 \tperp x/\vf )
\right. \nonumber \\ && \left.  \hspace{.5cm} 
      - \cos \sqrt{2} \theta_{{\rm C -}} 
      - \cos \sqrt{2} \theta_{{\rm S +}} 
      - \cos \sqrt{2} \theta_{{\rm S -}} 
    \right]            \nonumber \\
&&-
    \frac{g_{3}}{2\pi^2 \alpha^2} \int \dx
    \sin  \sqrt{2} \theta_{\rho +} 
    \left[
        \cos ( \sqrt{2}\theta_{{\rm C +}} - 8 \tperp x/\vf)
\right. \nonumber \\ && \left. \hspace{.5cm}
      + \cos \sqrt{2} \theta_{{\rm C -}} 
      - \cos \sqrt{2} \theta_{{\rm S +}} 
      + \cos \sqrt{2} \theta_{{\rm S -}} 
    \right]         
                                   \virg 
\label{phase_Hamiltonian}
\end{eqnarray}
 where 
$v_{\rho(\sigma)} = \vf [1+\!(-)U/\pi\vf]^{1/2}$,
$K_{\rho(\sigma)} =  [1+\!(-)Ua/\pi\vf]^{-1/2}$,
$v_{\rm C} =v_{\rm S} = \vf$, $K_{\rm C}= K_{\rm S} = 1$,
 $g_{\rho(\sigma)} = + (-) U a$ and 
    $g_3 = U a (2 \td/t ) /[1+(\td/t)^2]$. 
 The quantity $\alpha$ is 
 a  cutoff of the order of lattice constant.
In eq.~(\ref{phase_Hamiltonian}), 
 there are twelve nonlinear terms   
  rewritten as 
\begin{eqnarray}
             \label{coupling}
\frac{g_{\nu p,\nu' p'}}{2\pi^2 \alpha^2}
 \int dx \cos \sqrt{2} \psi_{\nu p} 
                    \cos \sqrt{2} \psi_{\nu' p'} \virg  
\end{eqnarray}
 where 
     $\psi_{\nu \pm} = \theta_{\nu \pm} $ 
       except for $\psi_{{\rm C} +} = 
        \theta_{\rm C +}-(8\tperp x/ \vf) /\sqrt{2}$ 
 and  
     $\psi_{{\rho} +} = 
        \theta_{\rho +} - \pi/(2\sqrt{2})$.
The RG equations for $ K_{\nu} = K_{\nu}(l), 
  t_{\perp}=t_{\perp}(l)$ and  
       $G_{\nu p,\nu' p'} = G_{\nu p,\nu' p'}(l)$  
    are given,  up to the second order, as  
  \cite{Suzumura, Tsuchiizu}
\begin{eqnarray}  
  \dl \tt_\perp &=& 
    \tt_\perp 
 -  \frac{1}{8}  \Kc \bigl( 
         \grcp^2 
\nonumber \\ && \hspace{.cm} 
                 +  \gpcp^2 + \gcpsp^2 + \gcpsm^2 
           \bigr)
         J_1(8\tt_\perp)
,
\label{eq:tperp} 
\\ 
\dl K_\nu &=& 
-\frac{1}{2\tilde{v}_\nu^2}  K_\nu^2
   \left[
    \gncp^2 
    J_0 ( 8\tt_\perp ) 
\right. \nonumber \\ && \hspace{0cm} \left.
   +  \gncm^2 + \gnsp^2 + \gnsm^2 
  \right] 
,
\label{eq:Knu}
\\ 
\dl \Kc &=& 
  - \frac{1}{2} \sum_{p=\pm} 
   \left[
     \left(\Kc^2  J_0( 8\tt_\perp)  \delta_{p,+} 
                                         - \delta_{p,-} \right)
       \left(\grc^2
\right. \right. \nonumber \\ && \hspace{.cm} \left. \left.
                    + \gpc^2 + \gcsp^2 + \gcsm^2 \right)
   \right] 
,
\label{eq:Kc}
\\
\dl \Ks &=&
 - \frac{1}{2} \sum_{p=\pm} 
   \left[
       \left( \Ks^2  \delta_{p,+} - \delta_{p,-} \right)
       \left( \grs^2 
\right. \right. \nonumber \\ && \hspace{.cm} \left. \left.
           + \gps^2 + \gcps^2  J_0( 8\tt_\perp )  + \gcms^2
       \right) 
   \right] 
,
\label{eq:Ks}
\\ 
\dl \gnc &=& 
    \bigl[ 2 - K_\nu - \Kc^{p} \bigr] \gnc
\nonumber \\ && \hspace{.cm} 
      - \gnsp \gcsp - \gnsm  \gcsm  , 
\label{eq:Gnu-c}
\\ 
\dl \gns &=& 
    \bigl[  2 - K_\nu - \Ks^{p} \bigr] \gns
\nonumber \\ && \hspace{.cm} 
      - \gncp \gcps  J_0( 8\tt_\perp )
\nonumber \\ && \hspace{.cm} 
      - \gncm \gcms ,  
\label{eq:Gnu-s}
\\ 
\dl \gcs &=& 
    \bigl[ 2 - \Kc^p - \Ks^{p'} \bigr] \gcs
\nonumber \\ && \hspace{.cm} 
    - \frac{1}{\tilde{v}_\rho} \grc  \grsd  
    - \frac{1}{\tilde{v}_\sigma} \gpc  \gpsd  , 
\label{eq:Gc-s}
\end{eqnarray}
where $\tilde{v}_\nu=v_\nu/\vf$ and  
 $\tt_\perp = t_{\perp}(l)/\vf \alpha^{-1}$.
 $J_n$ is the $n$-th Bessel function.   The quantity $l$ 
  is  related to energy scale $\omega$ or temperature $T$ by   
  $l=\ln(W/\omega)$ or $\ln(W/T)$ with $W(\equiv \vf \alpha^{-1})$ 
  being of the order of band width.  
The initial condition for the RG equations are given by 
  $K_\nu(0)=K_\nu$, 
   $G_{\nu p, \nu' p'}(0)= g_{\nu p, \nu' p'}/2\pi\vf$ and
  $\ttperp(0)= \tperp/(\vf \alpha^{-1})$ 
 where 
  ${g_{{\rm C} + ,{\rm S} +}} = -{g_{{\rm C} - ,{\rm S} -}} 
         = 0 $, 
  ${g_{{\rm C} + ,{\rm S} -}} = -{g_{{\rm C} - ,{\rm S} +}} 
         = - U a $, 
  ${g_{\sigma +, {\rm C} +}} =-{g_{\sigma +, {\rm C} -}} 
         = -{g_{\sigma +, {\rm S} +}} =-{g_{\sigma +, {\rm S} -}} 
         = U a$, 
 and 
  ${g_{\rho +, {\rm C} +}} = {g_{\rho +, {\rm C} -}} 
         = - {g_{\rho +, {\rm S} +}} = {g_{\rho +, {\rm S} -}} = g_3$.
 We take $\alpha=2a/\pi$ \cite{Tsuchiizu_SCES} 
and  discard the RG equations  for the velocity $v_\nu$.

\section{Confinement-deconfinement transition}
 
We   calculate  eqs.~(\ref{eq:tperp})-(\ref{eq:Gc-s}) 
 numerically  by choosing $U$, $\td$, and $t_{\perp}$ as parameters.  

\begin{figure}[b]
\begin{center}
\vspace*{10mm}
\leavevmode
\epsfysize=7.5cm
   \epsffile{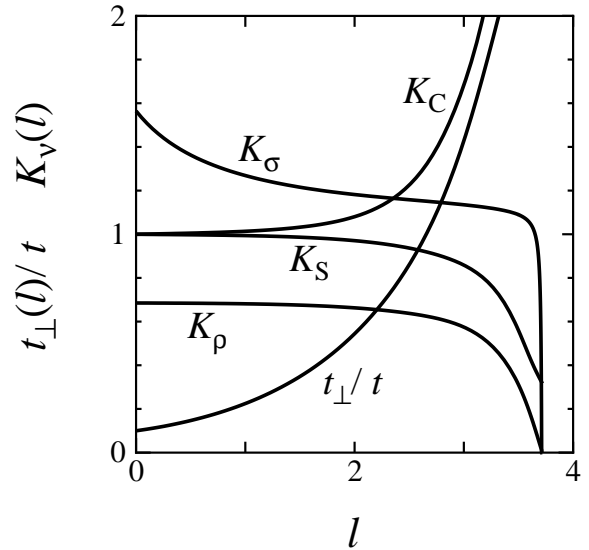}
\end{center}
\vspace*{-2mm}
\caption{
The $l$-dependence of $K_{\rho}(l)$, $K_{\sigma}(l)$, 
$\Kc(l)$, $\Ks(l)$ and $t_{\perp}(l)$ for 
 $U/t=5$, $t_{\perp}/t$ = 0.1 and $\td/t = 0.05$.
}
\vspace{-10mm}
\end{figure}
Figure 1 shows  the $l$-dependence of $K_{\rho}(l)$, $K_{\sigma}(l)$, 
  $\Kc(l)$, $\Ks(l)$ and $t_{\perp}(l)$ 
 which exhibit four  gaps. 
 With increasing $l$,
  $K_{\rho}(l)$ decreases to zero forming a  gap 
  in the total charge fluctuation 
  while  
  $\Kc(l)$ increases infinity to induce a gap in 
  the  transverse charge fluctuation.   
 Quantities $K_{\sigma}(l)$ and $\Ks(l)$ decrease also 
    to zero and lead to  spin gaps for both the total and transverse 
   spin fluctuations.  
 The rapid increase of $t_{\perp}(l)$ comes from a fact that 
   the term with $\Kc(l)$ in the r.h.s. of eq.~(\ref{eq:tperp}) 
     reduces to zero due to a factor $J_1(8\tilde{t}_{\perp})$.  
 Figure 2 displays the corresponding $l$-dependence of coupling constants. 
 The main figure shows  
  coupling constants for forward and backward scatterings  with 
   $G_{{\rm C}+,{\rm S}+}$ (curve (1)), 
               $- G_{{\rm C}+,{\rm S}-}$ (curve (2)),
    $G_{{\rm C}-,{\rm S}+}$ (curve (3)), 
       $-G_{{\rm C}-,{\rm S}-}$ (curve (4)),
     $G_{\sigma +,{\rm C}+}$ (curve (5)), 
            $- G_{\sigma +,{\rm C}-}$ (curve (6)), 
      $G_{\sigma +,{\rm S}+}$ (curve (7))  
     and  $-G_{\sigma +, {\rm S}-}$ (curve (8)) 
 while the inset shows those for the umklapp scattering with 
  $G_{\rho +,{\rm C}+}$ (curve (9)),  
        $G_{\rho +,{\rm C}-}$ (curve (10)),  
          $-G_{\rho +,{\rm S}+}$ (curve (11))   and   
         $G_{\rho +,{\rm S} -}$ (curve (12)).   
 Coupling constants 
 $G_{\rho +,{\rm C}-} $ and $-G_{\rho +,{\rm S}+} $
  increase  rapidly and give rise to the  trigger of relevance of 
     the  coupling constant $G_{{\rm C}-,{\rm S}+}$ 
       as seen also in the one-dimensional chain. 
 Note that the relevance  of   coupling constants  
   $G_{{\rm C}-,{\rm S}+}$,  $- G_{\sigma +,{\rm C}-}$  
    and  $-G_{\sigma +,{\rm S}+}$  
      is also obtained in the absence of umklapp scattering \cite{Fabrizio}. 
The relevant behaviors  found   from the zero limit of 
  $K_{\rho}$, $K_{\sigma}$, $1/\Kc$ and  $\Ks$  exhibit the phase locking of  
       $\theta_{\rho +}$,  $\theta_{\sigma +}$,  $\theta_{{\rm C} -}$ and
           $\theta_{{\rm S} +}$, which are given by    
$\sqrt{2} \theta_{\rho +} = \pi/2$, 
 $\sqrt{2}\theta_{\sigma +} = 0$,
 $\sqrt{2}\theta_{{\rm C} -} = 0$ and
 $\sqrt{2} \theta_{{\rm S} +} = \pi$  
  from relevant behaviors of curves (3), (6), (7), (10) and (11). 
\begin{figure}[t]
\begin{center}
\vspace*{6mm}
\leavevmode
\epsfysize=7.5cm
   \epsffile{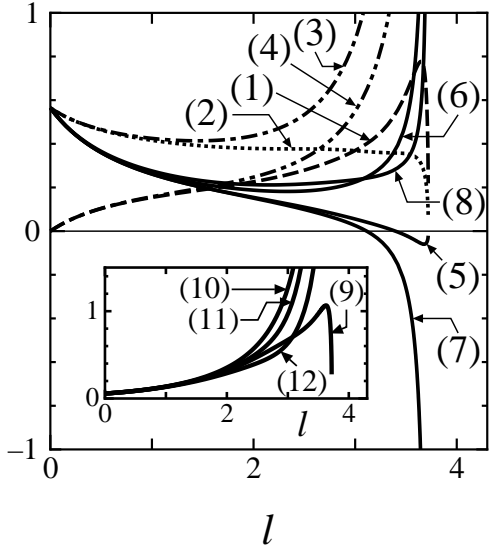}
\end{center}
\vspace*{-2mm}
\caption{
The $l$-dependence 
 of coupling constants $G_{\nu p, \nu' p'}(l)$
 for with  $U/t=5$, $t_{\perp}/t$ = 0.1 and $\td/t = 0.05$ 
 where curves (1)-(12) are explained in the text. 
}
\end{figure}
\begin{figure}[t]
\begin{center}
\vspace*{6mm}
\leavevmode
\epsfysize=7.5cm
   \epsffile{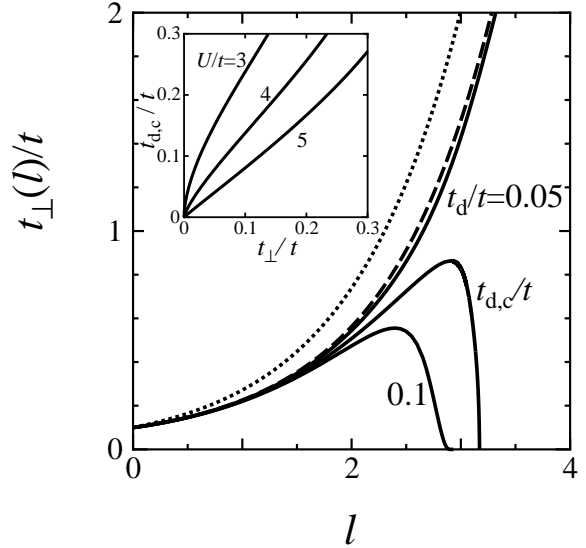}
\end{center}
\vspace*{-2mm}
\caption{
The $l$-dependence of $t_{\perp}(l)$  for 
 $U/t=5$ and $t_{\perp}/t=0.1$  with  
  the fixed $\td/t =$ 0.05, $t_{\rm d,c}/t(\simeq 0.082)$ and 0.1.  
 The dotted (dashed) curve shows  $t_{\perp}(l)$ 
        for $U$ =  0 
     ($U/t$ = 5, $\td/t = 0.05$  but $g_3$ = 0).  
 The inset denotes  
$t_{\rm d,c}$ 
 as the function of $t_{\perp}$ for the fixed   $U/t$ = 3, 4, and 5. 
}
\end{figure}
 Other coupling constants, 
  which are  expected to decrease \cite{Fabrizio},    
   are still large due to  
     the second order perturbation.   
 The change of the sign of   $G_{\sigma +, {\rm S}+}$  
  in the renormalization process comes from  
    the relevance of   $\theta_{\sigma +}$ and  $\theta_{{\rm S} +}$.
The effect of $t_{\perp}$ on coupling constants becomes large  
  at low energies 
 where the splitting of magnitudes becomes noticeable  for 
  the forward scattering 
 (between curves (1) and (4) 
  and between  curves (2) and (3)),
  the backward scattering (curves (5), (6), (7) and (8)),
 and the umklapp scattering  
 (curves (9), (10), (11) and (12)).

In Fig.~3, the $l$-dependence of $t_{\perp}$ is shown for 
  the fixed $\td/t =$ 0.05, $t_{\rm d,c}/t(\simeq 0.082)$ and 0.1. 
 The increase of $\td$ leads to the suppression of $t_{\perp}(l)$. 
   The case of $\td/t = 0.05$ shows the relevant behavior, which 
     corresponds to  deconfinement. 
 The quantity $t_{\perp}(l)$  for  $\td/t = 0.1$  does not increase 
   monotonically but decreases to zero after taking  a maximum  
    indicating    confinement. 
 The quantity  $t_{\perp}(l)$  with  a critical magnitude of 
   $\td = t_{\rm d,c}$ 
    denotes the behavior   
   between the confinement and the deconfinement. 
 For comparison, we show the dotted curve (the dashed) curve 
  which is  calculated for $U =  0$ 
     ($U/t$ = 5, $\td/t$ = 0.05  but $g_3$ = 0 
       as a special choice of parameter) 
  where  the analytical expression for the dotted curve is 
 given by 
 $t_{\perp}(l) = t_{\perp} \e^l$. 
 Note that the dashed curve is different from the case of 
    $U/t=5$ and $\td = 0$. 
 The dashed  curve 
   is evaluated to obtain  $t_{\perp}^{\eff0}$, which denotes 
    the interchain hopping renormalized only by the intrachain 
    interaction, i.e., without the umklapp scattering. 
 The effective interchain hopping $t_{\perp}^{\eff0}$ 
 is defined by $t_{\perp}^{\eff0} =  t \exp [- l_{\eff0}]$ 
   where  $t_{\perp}(l_{\eff0})/t = 1$ for $g_3=0$. 
 The  quantity $t_{\perp}^{\eff0}/t$  becomes unity for 
   $U$ = 0 or  $t_{\perp}/t$  = 1.
\begin{figure}[b]
\begin{center}
\vspace*{6mm}
\leavevmode
\epsfysize=7.5cm
   \epsffile{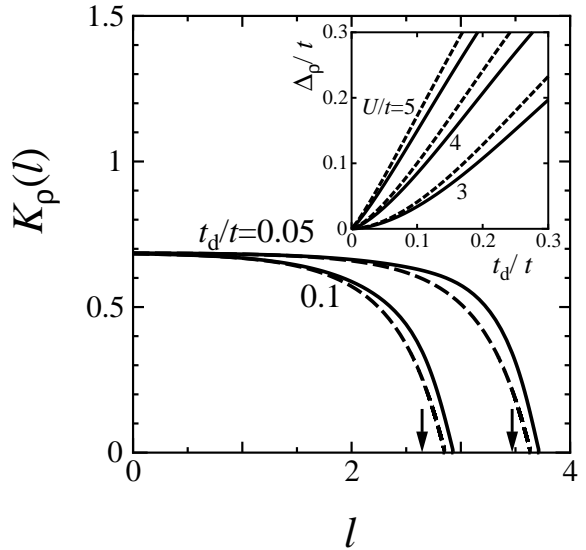}
\end{center}
\vspace*{-2mm}
\caption{
 The $l$-dependence of 
  $K_{\rho}(l)$  for  $U/t=5$  with  the fixed 
    $\td/t = 0.05$ and 0.1 where 
 $t_{\perp}$ = 0 for the dashed curve  and  $t_{\perp}$ = 0.1 for 
the solid curve. 
 The arrow denotes $l_{\Delta}$ defined by 
     $K_{\rho}(l_{\Delta}) = K_{\rho}/2$.
 The inset exhibits the charge gap: 
 $\Delta_{\rho}^{\rm 1D}$     for  $t_{\perp}/t$ = 0 (dashed curve) 
 and $\Delta_{\rho}$ for $t_{\perp}/t$ =  0.1 (solid curve). 
}
\vspace{-10mm}
\end{figure}
 The inset denotes the $t_{\perp}$-dependence 
  of $t_{\rm d,c}$ for the fixed 
    $U/t$ = 3, 4, and 5. 
 The boundary between confinement ($\td > t_{\rm d,c}$) 
  and deconfinement ($\td < t_{\rm d,c}$) depends  appreciably on $U$, 
    which gives rise to the enhancement of the confined region 
       on the plane of $t_{\perp}$ and $t_{\rm d,c}$.
 The limiting form for small $t_{\perp}$ 
        is given by 
        $t_{\rm d,c}/t \propto   (t_{\perp}/t)^{F}$ with 
                  the $U$-dependent $F$.

The $l$-dependence of $K_{\rho}(l)$ 
  is shown by solid  curve for the fixed 
    $\td/t$ = 0.05 and 0.1 in Fig.~4.
 The charge gap is defined by 
  $\Delta_{\rho} = \vf \alpha^{-1}  \exp [-l_{\Delta}]$ 
    with  $K_{\rho}(l_{\Delta}) = K_{\rho}/2$ and 
            $\vf \alpha^{-1} = t (\pi/\sqrt{2})
                  [1-(\td/t)^2]/[1+(\td/t)^2]^{1/2}$.  
 The dashed curve, which  denotes  $K_{\rho}(l)$ for $t_{\perp}=0$, 
  leads to  the charge gap, $\Delta_{\rho}^{\rm 1D}$,  
 for one-dimensional (1D) case 
  in the presence of dimerization, $\td$. 
 The charge gap is suppressed slightly by the interchain hopping
  since $K_{\rho}(l)$ for the solid curve   decreases slowly 
        compared with  that for the  dashed curve, i.e., 
       $l_{\Delta }$ is increased by $t_{\perp}$.  
 Such a behavior is understood from 
   eq.~(\ref{eq:Knu}) with $\nu=\rho$, 
         in which  the first term of the r.h.s.~becomes small 
            due to the  Bessel function suppressed 
                     by the large  $t_{\perp}$. 
 The inset exhibits the charge gap  
   as a function of $\td$ 
     with the fixed $U/t$ = 3, 4 and 5 where the solid (dashed) curve 
     corresponds to the case of $t_{\perp}/t$ = 0.1 (0). 
     Note that  the dashed curve is  given by 
$\Delta_{\rho}^{\rm 1D} \simeq W (g_3/W)^{1/(2-2K_\rho)}$ \cite{Schwartz}.

\begin{figure}[t]
\begin{center}
\vspace*{6mm}
\leavevmode
\epsfysize=7.5cm
   \epsffile{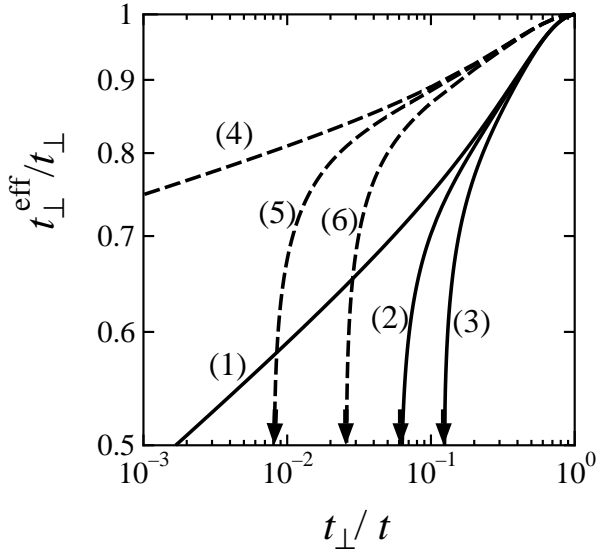}
\end{center}
\vspace*{-2mm}
\caption{
 The $t_{\perp}$-dependence of $t_{\perp}^{\rm  eff}$ 
  with $U/t$ = 3 (dashed curve) and 5 (solid curve) 
 where 
 $\td/t$ = 0 (curves (1) and (4)),  
  0.05 (curves (2) and (5)) and   
  0.1 (curves (3) and (6)).   
}
\end{figure}
In Fig.~5, 
 the $t_{\perp}$-dependence of $t_{\perp}^{\rm  eff}$ 
  with $U/t$ = 5  ($U/t$ = 3)
 is shown 
 by the solid (dotted) curve  for 
 $\td/t$ = 0 (1), 0.05 (2) and 0.1 (3)  
  ($\td/t$ = 0 (4), 0.05 (5) and 0.1 (6)), 
  where the effective interchain hopping $t_\perp^{\rm eff}$, 
  including the effect of the dimerization, is estimated by 
  $t_\perp^{\rm eff}=t \exp[-l_{\rm eff}]$ with 
  $t_\perp(l_{\rm eff})/t=1$ 
\cite{Tsuchiizu_PRG}.   
 The curves (1) and (4) for small $t_{\perp}/t$
 well reproduce the analytical 
     result given by
   $t_{\perp}^{\rm eff}/t_\perp \propto 
    (t_{\perp}/t)^{\alpha_0/(1-\alpha_0)}$ 
   with  the $U$-dependent quantity 
   $\alpha_0=(K_\rho+K_\rho^{-1}+K_\sigma+K_\sigma^{-1}-4)/4$
 \cite{Bourbonnais}.
 Comparing  the slope of  curve (1) with that of curve  (4), 
   one finds  that 
   the renormalization of $t_{\perp}$
      increases   by the intrachain interaction 
        $U$. 
With decreasing $t_{\perp}/t$, 
  the ratio $t_{\perp}^{\rm eff}/t_\perp$ decreases  rapidly 
    and becomes zero for 
  $t_{\perp}$ less than a critical value of $t_{\perp}$ 
    indicated by the arrow.

\begin{figure}[t]
\begin{center}
\vspace*{6mm}
\leavevmode
\epsfysize=7.5cm
   \epsffile{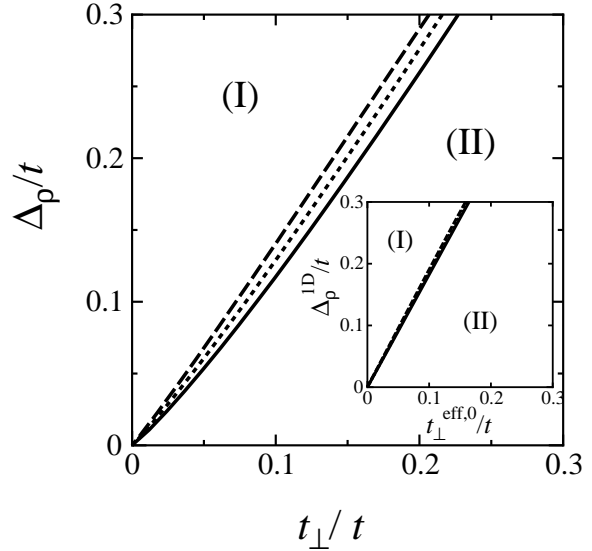}
\end{center}
\vspace*{-2mm}
\caption{
 The phase diagram of confinement (I) and deconfinement (II) on the 
   plane of $t_{\perp}$ and $\Delta_{\rho}$ 
 for 
    $U/t$ = 3 (dashed curve), 4 (dotted curve) and 5 (solid curve). 
 The inset denotes the corresponding phase diagram on the plane of 
     $t_{\perp}^{\eff0}$ and  $\Delta_{\rho}^{\rm 1D}$. 
}
\end{figure}
 From  the inset of Fig.~3 and   that  of Fig.~4, we obtain     
 the phase diagram of confinement (I) and deconfinement  (II) on the 
   plane of $t_{\perp}$ and $\Delta_{\rho}$.
  In Fig.~6, the boundary between these two states is shown for 
    $U/t$ = 3 (dashed curve), 4 (dotted curve) and 5 (solid curve). 
  The boundary is  rather straight and 
    the  $U$-dependence  becomes small 
      compared with that of the inset of Fig.~3.     
  It has been claimed previously that  
   such a behavior indicates  the  competition between 
    the charge  gap and the interchain hopping $t_{\perp}$
  \cite{Suzumura,Tsuchiizu,Tsuchiizu_SCES}.
  However such a statement is not clear enough 
  since the boundaries depend on the choice of $U$.
This problem can be remedied by the following treatment.
 We  take $\Delta_{\rho}^{\rm 1D}$  ($t_{\perp}^{\eff0}$)
          in stead of  $\Delta_{\rho}$  ($t_{\perp}$)
   as the vertical (horizontal) axis on the phase diagram 
  where  $\Delta_{\rho}^{\rm 1D}$ is obtained from the dashed curve of
 the inset of  Fig.~4. 
 The resultant boundaries are shown in the inset of Fig.~6, 
    where the good coincidence is obtained among   these three curves. 
  Thus it is concluded that the boundary between 
    confinement and deconfinement is determined by the competition 
     between the one-dimensional charge gap 
   (i.e., in the absence of the interchain hopping) and the 
    interchain hopping renormalized only  
         by the intrachain interaction (i.e., 
 without the umklapp scattering).

\section{Discussion} 
We have examined the mechanism of confinement 
    in terms of quarter-filled two-coupled chains with 
 dimerization 
 as a model of 
  low dimensional systems. 
 The confinement occurs when the charge gap induced by the umklapp
   scattering 
    becomes larger than the interchain hopping renormalized 
       by the intrachain hopping.
  It has been found that the ratio for   $U/t = 5$ ($U/t=3$) 
     and $ 0.05 <t_{\perp}/t < 0.2$ 
 is given by 
   $\Delta_{\rho}^{\rm 1D}/ t_{\perp}^{\eff0}  \simeq  1.8$ 
   ($\Delta_{\rho}^{\rm 1D}/ t_{\perp}^{\eff0}  \simeq  1.9$)
 and       
     $ 1.1 \lsim \Delta_{\rho}/ t_{\perp} \lsim 1.3$ 
     ($ 1.3 \lsim \Delta_{\rho}/ t_{\perp} \lsim 1.5$).

\begin{figure}[t]
\begin{center}
\vspace*{6mm}
\leavevmode
\epsfysize=7.5cm
   \epsffile{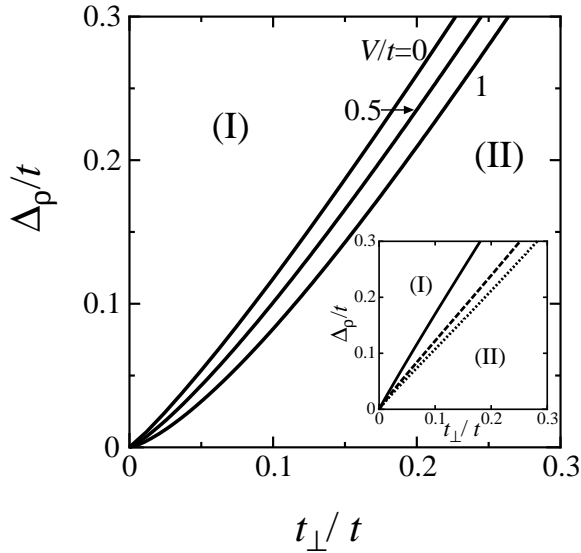}
\end{center}
\vspace*{-2mm}
\caption{
 The phase diagram of confinement (I) and deconfinement (II) 
    on the  plane of $t_{\perp}$ and $\Delta_{\rho}$. 
 The main figure denotes the boundary  for $U/t$ = 5 and $\td/t= 0.05$
       with the fixed nearest-neighbor interaction:  
  $V/t$ = 0, 0.5 and 1. 
 In the inset,  
  the dotted (dashed) curve  
   denotes  the boundary for   
        the fixed $\td/t$ = 0.05 (0.1)
 while the solid curve denotes 
     the boundary for the another case of the Hubbard model 
  at half-filling with $\td/t=0$. 
}
\end{figure}
 Here we discuss the effect of nearest neighbor interaction ($V$),   
   for the small $V$, in which 
   the commensurability energy  may be  negligibly small
  \cite{Mila,Yoshioka}. 
In this case,  $V$ does not contribute 
   to  both the backward scattering  and the umklapp scattering 
        even  in the presence of the dimerization ($\td$).  
 The effect of  $V$ appears in  the  forward scattering 
 where the coupling constant is 
   expressed as 
  $g_{2^\para} = g_{4^\para} =2Va$ and    
  $g_{2^\perp}= g_{4^\perp} = (U+2V)a $. 
In Fig.~7,      
 the  boundary of confinement-deconfinement transition with $U/t=5$
is shown for 
 $V/t$ = 0, 0.5  and 1 
where $V$ enhances the confined region.

Finally we comment on the case of two-coupled chain with  
 the conventional half-filled Hubbard model where 
 the phase diagram   is shown in the inset of Fig.~7. 
 Compared with  those of Fig.~6, 
  the  ratio   $\Delta_{\rho}/t_{\perp}$ 
   ( $\simeq 1.7$ for  $t_{\perp}/t = 0.1$) 
      is slightly large  and the boundary (solid curve)
 is rather straight.
  Such a  result is compared with 
  the boundaries for 
    the fixed $\td/t$ = 0.05 and  0.1, which   
 are shown by the dotted  curve and dashed curve,  respectively. 
 With increasing  the dimerization,   
   the slope of the boundary becomes steep  
     and moves toward  the solid curve 
  since the large dimerization may lead  the system to   
   half-filling.

\section*{ Acknowledgements }

This work was supported by a Grant-in-Aid for Scientific 
  Research from the Ministry of Education, Science, Sports and
  Culture (Grant No.09640429), Japan.

\end{document}